\documentclass[iop]{emulateapj}

\usepackage{graphicx}
\usepackage{amsmath}
\usepackage{subfigure}
\usepackage{float}
\usepackage{natbib} 
\usepackage{hyperref}
\usepackage{tabularx}

\begin{document}
\title{The First Naked-Eye Superflare Detected from Proxima Centauri}

\author{Ward S. Howard\altaffilmark{1}, Matt A. Tilley\altaffilmark{2}, Hank Corbett\altaffilmark{1}, Allison Youngblood\altaffilmark{3}, R. O. Parke Loyd\altaffilmark{4}, Jeffrey K. Ratzloff\altaffilmark{1}, Nicholas M. Law\altaffilmark{1}, Octavi Fors\altaffilmark{1,5}, Daniel del Ser\altaffilmark{1,5}, Evgenya L. Shkolnik\altaffilmark{4}, Carl Ziegler\altaffilmark{1}, Erin E. Goeke\altaffilmark{1}, Aaron D. Pietraallo\altaffilmark{1}, Joshua Haislip\altaffilmark{1}}
\altaffiltext{1}{Department of Physics and Astronomy, University of North Carolina at Chapel Hill, Chapel Hill, NC 27599-3255, USA}
\altaffiltext{2}{Dept. of Earth \& Space Sciences, University of Washington, Seattle, WA, USA}
\altaffiltext{3}{NASA Goddard Space Flight Center, Greenbelt, MD, 20771, USA}
\altaffiltext{4}{School of Earth and Space Exploration, Arizona State University, Tempe, AZ, 85282, USA}
\altaffiltext{5}{Institut de Ci\`encies del Cosmos (ICCUB), Universitat de Barcelona, IEEC-UB, Mart\'{\i} i Franqu\`es 1, E08028 Barcelona, Spain}

\email[$\star$~E-mail:~]{wshoward@unc.edu}

\begin{abstract}
Proxima b is a terrestrial-mass planet in the habitable-zone of Proxima Centauri. Proxima Centauri's high stellar activity however casts doubt on the habitability of Proxima b: sufficiently bright and frequent flares and any associated proton events may destroy the planet's ozone layer, allowing lethal levels of UV flux to reach its surface. In March 2016, the Evryscope observed the first naked-eye-brightness superflare detected from Proxima Centauri. Proxima increased in optical flux by a factor of $\sim$68 during the superflare and released a bolometric energy of $10^{33.5}$ erg, $\sim$10$\times$ larger than any previously-detected flare from Proxima. Over the last two years the Evryscope has recorded 23 other large Proxima flares ranging in bolometric energy from $10^{30.6}$ erg to $10^{32.4}$ erg; coupling those rates with the single superflare detection, we predict at least five superflares occur each year. Simultaneous high-resolution HARPS spectroscopy during the Evryscope superflare constrains the superflare's UV spectrum and any associated coronal mass ejections. We use these results and the Evryscope flare rates to model the photochemical effects of N$\mathrm{O}_x$ atmospheric species generated by particle events from this extreme stellar activity, and show that the repeated flaring may be sufficient to reduce the ozone of an Earth-like atmosphere by 90\% within five years; complete depletion may occur within several hundred kyr. The UV light produced by the Evryscope superflare would therefore have reached the surface with $\sim$100$\times$ the intensity required to kill simple UV-hardy  microorganisms, suggesting that life would have to undergo extreme adaptations to survive in the surface areas of Proxima b exposed to these flares.
\end{abstract}


\maketitle

\section{Introduction}

The small and cool star Proxima Centauri (hereafter Proxima) hosts Proxima b, a likely rocky planet \citep{anglada2016,bixel2017} within the habitable zone (e.g. \citealt{Ribas2016,Meadows2018}). Proxima b has potential difficulties in maintaining a habitable atmosphere, both due to possible tidal locking \citep{griessmeier2009} and incident stellar activity (e.g. \citealt{tarter2007,scalo2007,seager2010,shields2016,davenport2016}).

Proxima is well-known to exhibit large stellar variability and to produce bright flare events. It is hypothesized that it can produce superflares, extreme stellar events with an estimated bolometric energy release of at least $10^{33}$ erg \citep{Segura2010,Tilley2017,loyd2018}; if detected, they would be one of the largest potential threats to the habitability of Proxima b \citep{davenport2016}: while ozone in an Earth-like planet's atmosphere can shield the planet from the intense UV flux associated with a single superflare \citep{Segura2010,Greissmeier2016,Tabataba2016}, the atmospheric ozone recovery time after a superflare is on the order of years \citep{Tilley2017}. A sufficiently high flare rate can therefore permanently prevent the formation of a protective ozone layer, leading to UV radiation levels on the surface which are beyond what some of the hardiest-known organisms can survive \citep{Greissmeier2016}. The cumulative effect of X-ray and UV flux from both stellar flares and quiescent emission can even strip planetary atmospheres over the course of several Gyr \citep{cuntz2016,owen2016}.

Many previous studies have explored low- and moderate-energy flare events on Proxima.  Optical surveys have found events with detected energies up to $10^{31.5}$ erg ($10^{32}$ erg bolometric) in visible light \citep{davenport2016}. ALMA recently detected a large sub-mm flare ($10^{28}$ erg in ALMA's Band 6), although multiwavelength flare studies are needed to determine how large sub-mm flares relate to flares in other bands and their habitability effects \citep{macgregor2018}. In the X-ray, events up to $10^{32}$ erg ($10^{32.7}$ erg bolometric) have been detected \citep{gudel2004}.


The MOST satellite \citep{Walker2003} performed the most comprehensive previous measurement of the Proxima flare rate. MOST observed Proxima for 37.6 days, observing 66 white-light flare events, the largest of which was $10^{31.5}$ erg in the MOST band-pass ($\sim$4500--7500 {\AA} ). No superflares were observed; extrapolating the cumulative flare frequency distribution (FFD) obtained by \citet{davenport2016} from the MOST flare sample out by 1.5 dex predicts $\sim$8 $10^{33+}$ erg events in the MOST bandpass occur per year.

To search for superflares and other short-timescale phenomena, the Evryscope \citep{Evryscope2015} is performing long-term high-cadence monitoring of Proxima, along with every other bright star in the Southern sky. In March 2016 the Evryscope detected the first-known Proxima superflare. Although no M-dwarfs are usually visible to the naked-eye \citep[e.g.,][]{shields2016}, Proxima briefly became at least a magnitude-6.8 star during this superflare, at the limit of visiblity to dark-site naked-eye observers.


\begin{figure}
\centering
\includegraphics[width=\columnwidth]{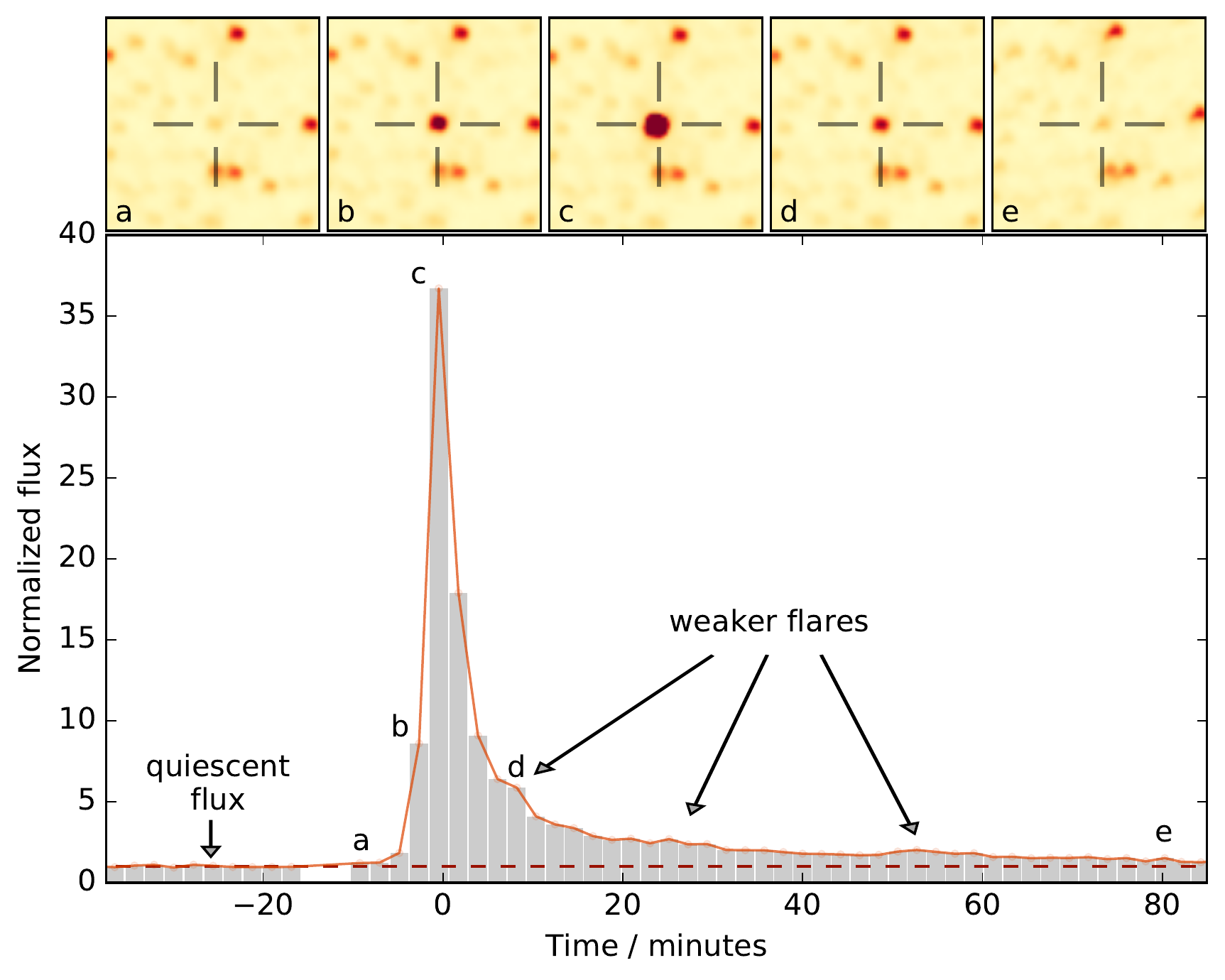}
\caption{The Evryscope discovery of a naked-eye-brightness superflare from Proxima. The y-axis is the flux increase over Proxima's median \textit{g'}-band flux from the previous hour. Bars show the integration time of each individual flux measurement. Insets display cutout images over the course of the flare. For clarity, we here show only one camera's light curve; another Evryscope camera simultaneously observing the event showed a very similar light curve offset by 2.2 seconds.
\label{fig:flare_event}}
\end{figure}

\section{Evryscope Flare Discovery and Observations}\label{evryscope_observations} \label{flare_det}

We discovered the Proxima superflare as part of the Evryscope survey of all bright Southern stars. The Evryscope is an array of small telescopes simultaneously imaging 8000 square degrees of the sky every two minutes~\citep{Evryscope2015}. The Evryscope observes essentially the entire Southern sky above an airmass of two, at two-minute cadence in \textit{g}\textsuperscript{$\prime$} and at a resolution of 13\arcsec pixel$^{-1}$. The system has a typical dark-sky limiting magnitude of \textit{g}\textsuperscript{$\prime$}=16 and tracks the sky for 2 hours at a time before ratcheting back and continuing observations, for an average of $\sim$6 hours of continuous monitoring each night on each part of the sky.

The Evryscope image archive contains 2.5 million raw images, $\sim$350TB of data. A custom pipeline analyzes the Evryscope dataset at realtime rates \citep{Law2016}. Each image, consisting of a 30 MPix FITS file from one camera, is calibrated using a custom wide-field solver. After careful background modeling and subtraction, forced-aperture photometry is extracted based on known source positions in a reference catalog. Light curves are then generated for approximately 15 million sources across the Southern sky by differential photometry in small sky regions using carefully-selected reference stars and a range of apertures; residual systematics are removed using two iterations of the SysRem detrending algorithm \citep{tamuz2005}. In extremely crowded fields, such as for Proxima ($-2^{\circ}$ galactic latitude), we re-run the pipeline for particular targets, optimizing the aperture sizes to avoid nearby stars.

As a very large event, the Proxima Superflare was discovered in routine by-eye checks of interesting targets in the Evryscope data set. Smaller flares are discovered and characterized with an automated flare-analysis pipeline which uses a custom flare-search algorithm, including injection tests to measure the flare recovery rate. First, we search for flares by attempting to fit an exponential-decay matched-filter similar to that of \citet{Liang2016} to each contiguous segment of the Evryscope light curve. Matches with a significance greater than 2.5$\sigma$ are verified by eye. The entire Proxima lightcurve is visually examined for flares to account for false-negatives in the automated search.

The fractional flux and equivalent duration (ED) for each flare are calculated as described in \citet{hawley2014}, and flare start and stop times are initially chosen where the flare candidate exceeds the local noise and are subsequently confirmed or adjusted by eye. We inject simulated flares separately into each light curve and perform 20 trials of randomly-located flare injection and attempted recovery per contiguous lightcurve segment. We average the results across the lightcurve to measure recovery completeness as a global function of flare contrast and ED and to quantify the error in contrast and ED for each flare.

\subsection{Simultaneous high-resolution spectra from HARPS}\label{HARPS_spectra}
The superflare reported here occurred during the three-month \emph{Pale Red Dot} campaign, which first revealed the presence of Proxima b \citep{anglada2016} using the HARPS spectrograph on the ESO 3.6m at La Silla, Chile \citep{Mayor2003}. The HARPS spectrum was taken at 2016 March 18 8:59 UT, 27 minutes after the flare peak at 8:32 UT. This single 1200 second exposure captured the majority of the flare tail, including $20\%$ of the total radiated flux. 

\section{Proxima Superflare Properties}\label{naked_eye_superflare}
The Evryscope detected the Proxima Superflare on 2016 March 18, 8:32:10 UT (MJD 57465.35568, see Figure \ref{fig:flare_event}). The flare lasted approximately one hour. The flare energy release was dominated by a single large event but subsequently showed a complex morphology, with three weaker flares (each more than doubling Proxima's \textit{g}\textsuperscript{$\prime$} -band brightness) observed.
 
\begin{figure*}
	\centering
	{
		\includegraphics[trim= 7 1 3 1,clip, width=6.9in]{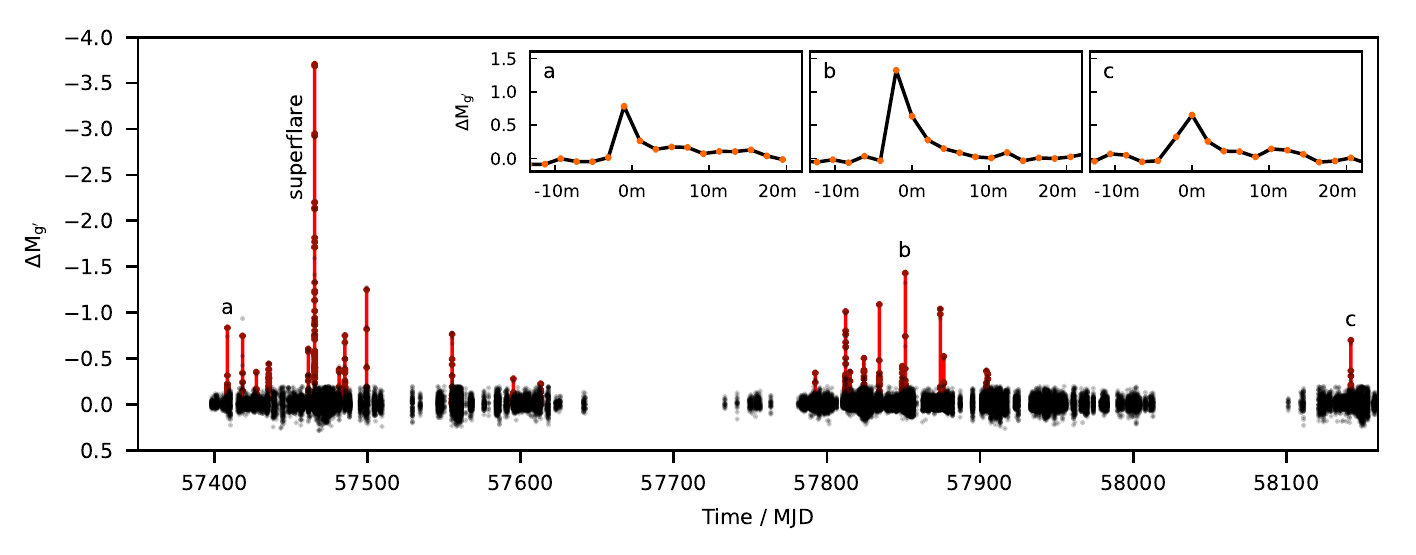}
	}
	\caption{The full 2016-18 Evryscope Proxima light curve. Detected flares are highlighted in red; to show short-term activity Proxima's long-term variability has been removed. The superflare is 2.5-magnitudes brighter than any other Evryscope-detected flare from Proxima. For clarity, we plot only 20\% of the 40,486 light curve points. Three representative flares are shown in detail. Because of the two-minute sampling, low-energy, short flares such as the rightmost individually-displayed flare often do not show classical rapid-rise flare shapes, although these flares are often confirmed by multiple cameras simultaneously.}
	\label{fig:full_lc}
\end{figure*}

\subsection{Peak brightness}\label{peak_bright} 
Within the Evryscope's two-minute integration during the flare peak, Proxima reached an average flux of 38$\times$ the quiescent emission. By fitting a instantaneous-flux flare template \citep{davenport2014} we estimate the superflare's brightness on human-eye timescales to have reached 68$\times$ Proxima's flux. Proxima, an 11.4 $g'$ magnitude star, thus briefly became a $g'$=6.8 star, at the limit of visibility to trained observers at very dark sites \citep{Weaver1947,Schaefer1990,Cinzano2001}.

\subsection{Energy release and planetary-impact-relevant fluxes}\label{total_energy}
Calculation of the superflare's atmospheric impacts requires an estimate of the flare's energy in multiple bandpasses, from the far-UV to the infrared. We measure the superflare energy in the \textit{g}\textsuperscript{$\prime$} Evryscope bandpass and subsequently convert into the bolometric energy using the energy partitions of \citet{Osten2015}. We accomplish this by estimating the bolometric flare energy of a 9000 K flare blackbody with emission matching the measured Evryscope flux; the fraction of the bolometric energy found in the Evryscope \textit{g}\textsuperscript{$\prime$}  bandpass is $f_{g'}$=0.19. The canonical value of 9000 K provides a lower limit to the flare energy; a higher-temperature flare blackbody, as has been sometimes measured for larger flares \citep{kowalski2010} results in more short-wavelength energy. The energy seen in any bandpass $\Delta \lambda$ is then given by the approximate relationship $E_{\Delta \lambda} = f_{\Delta \lambda} \times E_\mathrm{bol}$. 

We obtain the quiescent flux in the Evryscope \textit{g}\textsuperscript{$\prime$} bandpass by scaling directly from the Evryscope-measured calibrated magnitude, and by weighting the flux-calibrated spectrum of Proxima used in \citet{davenport2016} by the Evryscope response function and scaling using Proxima's distance \citep{lurie2014}. Both methods measure Proxima's quiescent flux in the Evryscope bandpass to be $L_0 = 10^{28.6}$ erg s$^{-1}$, giving the superflare energy in the Evryscope bandpass of $10^{32.8}$ erg, and a bolometric energy of $10^{33.5}$ erg.

\subsection{Proxima's flare frequency distribution}\label{evry_flare}

The Evryscope observed Proxima for a total of 1344 hours between January 2016 and March 2018. We discovered 24 large flares (Figure \ref{fig:full_lc}), with bolometric energies from $10^{30.6}$ erg to $10^{33.5}$ erg. To obtain the cumulative flare frequency distribution (FFD), we calculate the uncertainty in the cumulative occurrence rate for each Evryscope flare with a binomial $1\sigma$ confidence interval statistic (following \citealt{davenport2016}). Errors in energy for high-energy flares are calculated using the inverse significance of detection; low-significance flares use the injection-and-recovery error estimate instead, to account for the possibility of correlated noise introducing bias. 

\begin{figure}[tpb]
	\centering
	{
		\includegraphics[trim= 10 13 1 0,clip, width=3.4in]{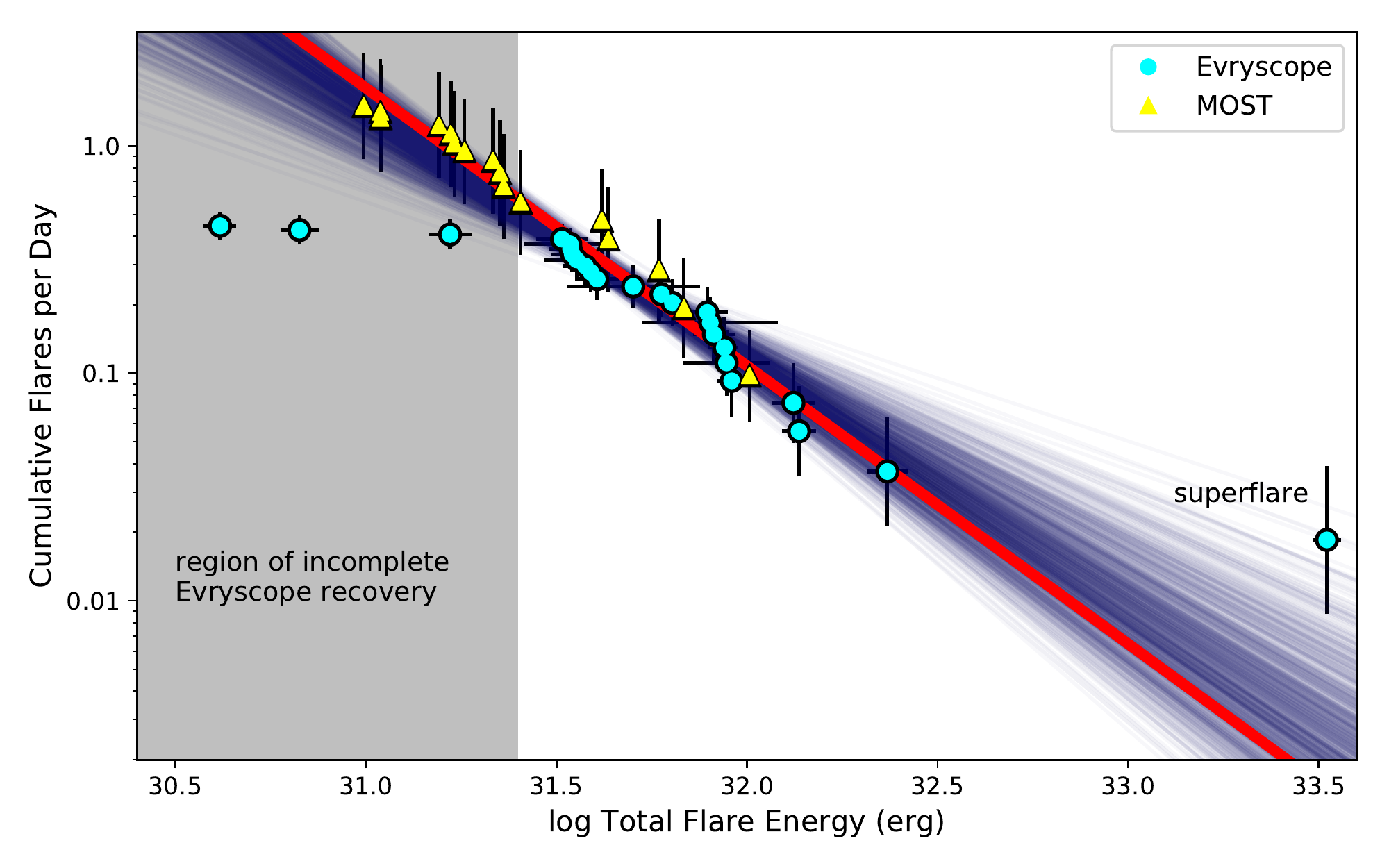}
	}
	\caption{The cumulative flare frequency distribution of Proxima fit to all Evryscope flares and the largest MOST flares, scaled to bolometric energy from the \textit{g}\textsuperscript{$\prime$} and MOST bandpass, respectively. The best fit, which excludes the Evryscope superflare, is displayed in red. 10,000 posterior draws (1000 shown) estimate the error of this power-law fit.}
	\label{fig:cffd}
\end{figure}

To sample both the rare high-energy events detectable by Evryscope and the frequent moderate-to-low energy events detectable by MOST, we also include flares from the MOST sample \citep{davenport2016} with energy in the MOST bandpass greater than $10^{30.5}$ erg (we exclude lower-energy MOST flares due to a possible knee in the FFD biasing the occurrence of superflares $\sim$1000X larger).  We fit a cumulative power-law FFD to the MOST and Evryscope flares, and determine the uncertainty in our fit through 10,000 Monte-Carlo posterior draws consistent with our uncertainties in energy and occurrence rates. We represent the cumulative FFD in the Evryscope bandpass (Figure \ref{fig:cffd}) by a power law of the form $\log{\nu}=(1-\alpha)\log{E} + b$, where $\nu$ is the number of flares with an energy greater than or equal to $E$ erg per day, $\alpha$ gives the frequency at which flares of various energies occur, and $b$ is the y-intercept and crossover into the unphysical energy region $E<0$.

Evryscope constrains the expected occurrence of $10^{33}$ erg bolometric events to be at least $5.2^{+0.2}_{-3.0}$ per year. It is evident from Figure \ref{fig:cffd} that it is difficult to fit a single power law which reproduces both the lower-energy flares and the Evryscope-observed superflare. This could mean the probability of reaching superflare energies is higher than would be expected by a simple power-law extrapolation from lower energies; it could also be that the Proxima Superflare is just a statistically-rare event. We therefore report two separate FFDs; the first excludes the Proxima Superflare, while the second includes it.  For the no-superflare case, we report an FFD of $\log{\nu} = \mathrm{-1.22}^{+0.26}_{-0.003} \log{E} + \mathrm{38.1}^{+8.4}_{-0.07}$, displayed in Figure \ref{fig:cffd}, in good agreement with both the Evryscope and MOST sample. Including the prior of the observed superflare, we obtain $\log{\nu} = \mathrm{-0.98}^{+0.02}_{-0.24} \log{E} + \mathrm{30.6}^{+0.83}_{-7.6}$. We note $\alpha_\mathrm{Evryscope}$ is significantly steeper in our higher-energy flare sample in both cases than that for Proxima FFDs from previous studies, e.g. \citet{Walker1981, davenport2016}.

\subsection{High-resolution flare spectrum}

\begin{figure}
	\centering
	{
		\includegraphics[width=3.4in]{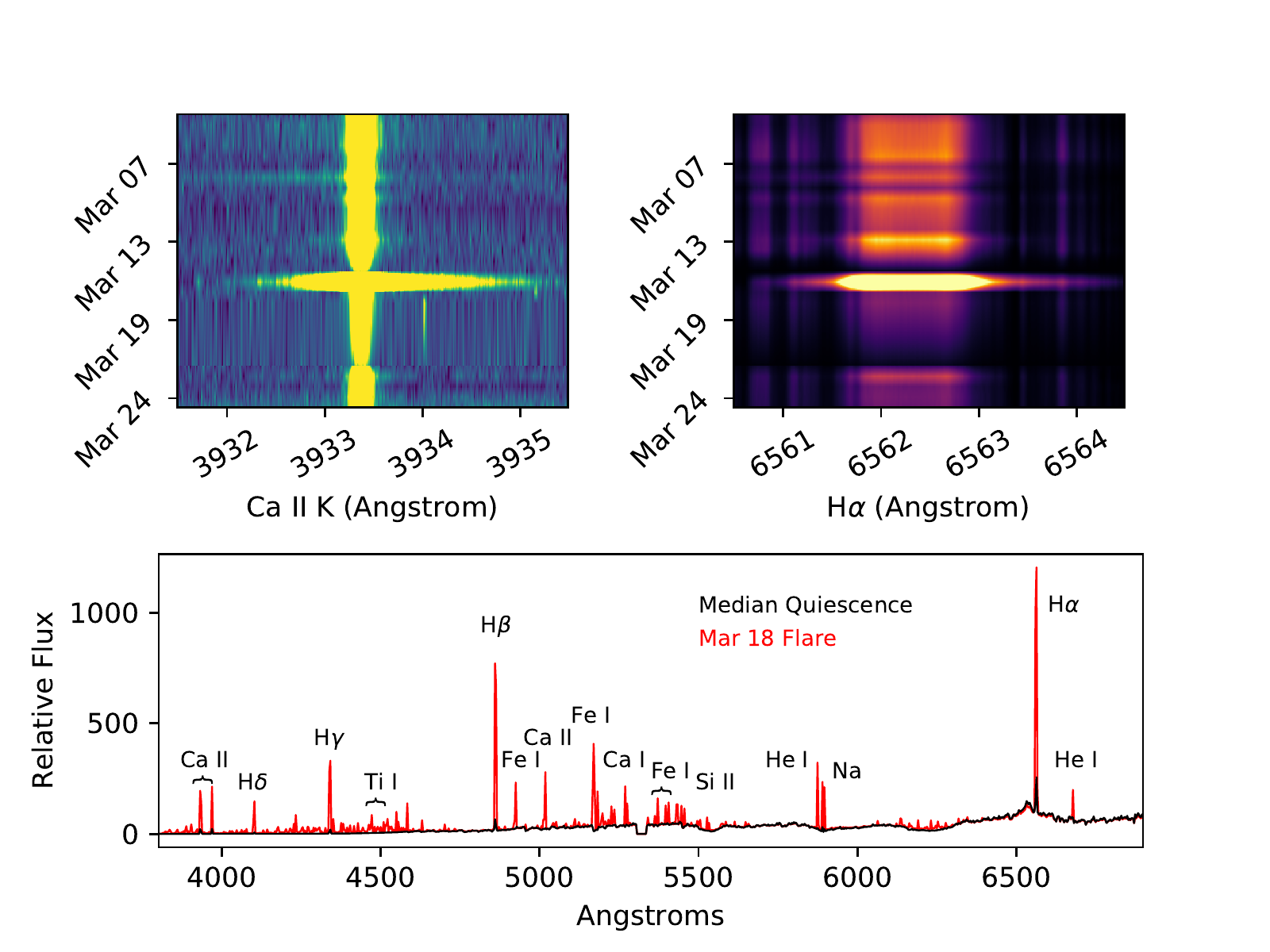}
		\label{fig:harps_proxima}
	}
	\caption{\emph{Top:} Chromospheric activity evolution traced by the Ca II (K) and $\mathrm{H}_\alpha$ indicators in the month leading up to the flare. \emph{Bottom:} Flux-normalized median quiescent spectrum from the month leading up to the flare (black) and active (red) spectrum 27 minutes after the superflare peak. Flare spectrum is relative to the normalized quiescent spectrum.}
  \label{fig:HARPS_combined}
\end{figure}

The spectrum of Proxima, in both the median quiescent and flare states, is shown in Figure~\ref{fig:HARPS_combined}. During the superflare, chromospheric metals and the Balmer series show sharply increased emission. A $-30$ km s$^{-1}$ splitting of the $\mathrm{H}_\alpha$, $\mathrm{H}_\beta$, and He I lines is detectable, and is indicative of a flow of highly ionized plasma generated by the flare, most likely correlated to a hot stellar wind moving outwards from the star \citep{Pavlenko2017}.  No significantly-blueshifted emission or anomalous emission lines are visible; the superflare spectrum is similar to other smaller flares recorded from Proxima and is therefore likely to be amenable to emission-line scaling relations to estimate Far-UV (FUV) and particle fluxes (Section \ref{sec:multiwavelength}).

\subsection{UV and particle fluxes}
\label{sec:multiwavelength}
FUV (912--1700 {\AA}) photons are capable of photolyzing most molecules in planetary atmospheres. The \textit{Hubble Space Telescope} archive contains 13.3~h of FUV spectrophotometry by the STIS spectrograph of Proxima. From this data, we aggregated 9 flares spanning FUV energies of $10^{29.3}$~--~$10^{30.8}$ erg to construct a FUV energy budget for Proxima flares. We scale to a 9000 K blackbody and EUV emission via a ``fiducial flare'' prescription \citep{loyd2018}, tailored to the Proxima data.  The $10^{32.5}$ erg FUV energy of the Proxima Superflare obtained by the fiducial flare prescription is found to be consistent with an independently-obtained estimate using the quiescent scaling relations of \cite{Youngblood2017} applied to a measurement of the Ca II K equivalent width in the HARPS spectrum (Section \ref{HARPS_spectra}).

\par Coronal mass ejections (CMEs) are often assumed to accompany large flares \citep{Yashiro2006}. \cite{Youngblood2017} measures a relation for predicting $>$10 MeV proton fluxes based on the energy of the flare in Si IV. These particles can initiate nonthermal chemical reactions in the planetary atmosphere. From the \cite{Youngblood2017} relation and the HARPS spectrum, we estimate a proton fluence at Proxima b's 0.0485 AU distance of $10^{7.7}$ protons cm$^{-2}$. 

\section{Astrobiological Impact of the Superflare }\label{planet_habitability}

\subsection{Demise of the Ozone Column}\label{ozone}
We employ a 1D coupled, photochemical and radiative-convective climate model to determine the effects of the observed flare activity on the potential habitability of Proxima b. We assume the planet to have an Earth-like atmosphere, but neglect the planetary magnetic field, which may be weaker than Earth's due to tidal-locking \citep{griessmeier2009}. The details of the coupled model can be found in \citet{Segura2010} and \citet{Tilley2017}, which discovered that the results of only electromagnetic flaring cannot significantly drive $\mathrm{O}_3$ column loss, while flares with proton events can rapidly destroy the $\mathrm{O}_3$ column. Proton events lead to the dissociation of $\mathrm{N}_2$ in the planet atmosphere into constituent, excited N-atoms, which then react with $\mathrm{O}_2$ to produce NO and O. NO reacts with $\mathrm{O}_3$ to produce $\mathrm{NO}_2$. The $\mathrm{NO}_x$ species generated during the proton events therefore drive the evolution of the ozone column (see \citealt{Tilley2017} for further details).

Using the cumulative FFD measured in the present work and the scaling from \citet{hawley2014}, we generate a sequence of flares for a 5-year time span in the U-band energy range of $10^{29.5}$ erg to $10^{32.9}$ erg (scaled to represent the Evryscope-measured FFD). We assume a time-resolved UV superflare spectrum scaled from AD Leo to Proxima flares, following \citet{Tilley2017}. This flare sequence drives the evolution of volatiles in an Earth-like atmosphere at a distance of 0.0485 AU to an active M dwarf. Flares are selected at random to produce a proton event, with proton flux scaling with event amplitude. The probability for each flare to generate a planet-oriented energetic particle event was assumed to be a moderate P = 0.08, following \citet{Tilley2017}.

\begin{figure}[t]
	\centering
    \subfigure
	{
		\includegraphics[trim= 60 60 210 525, clip, width=3.4in]{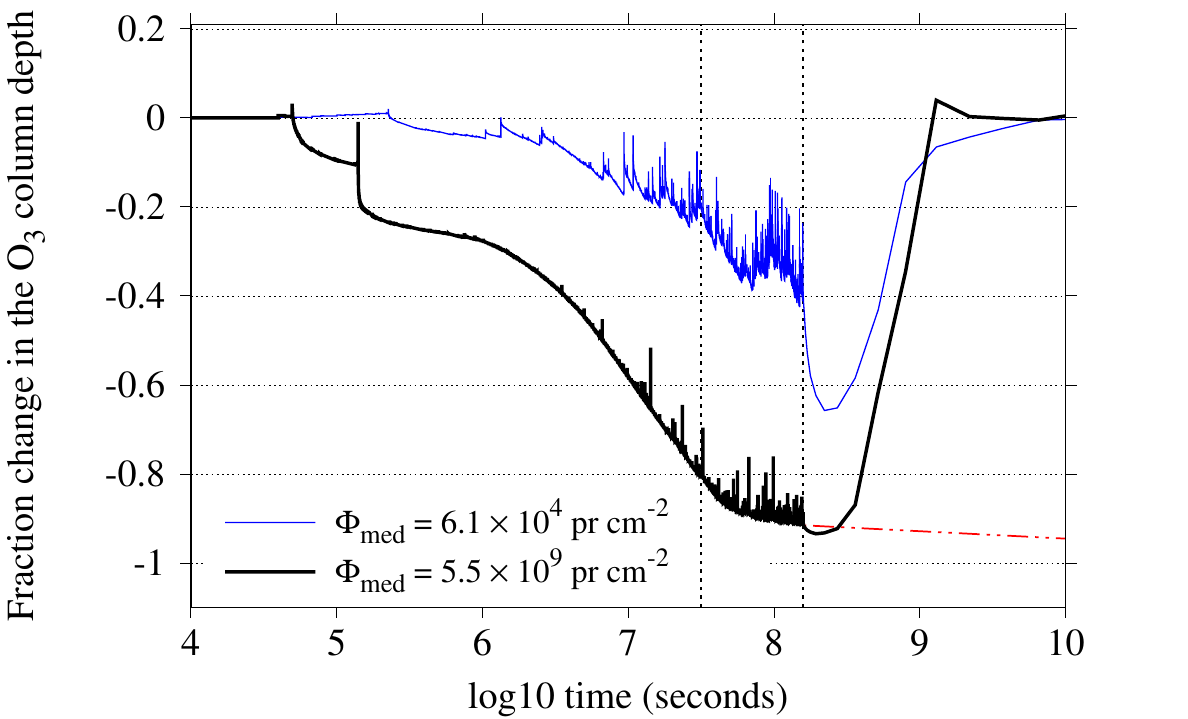}
		\label{fig:auto_vars}
	}
	\subfigure
	{
		\includegraphics[trim= -25 10 1 40, clip, width=3.4in]{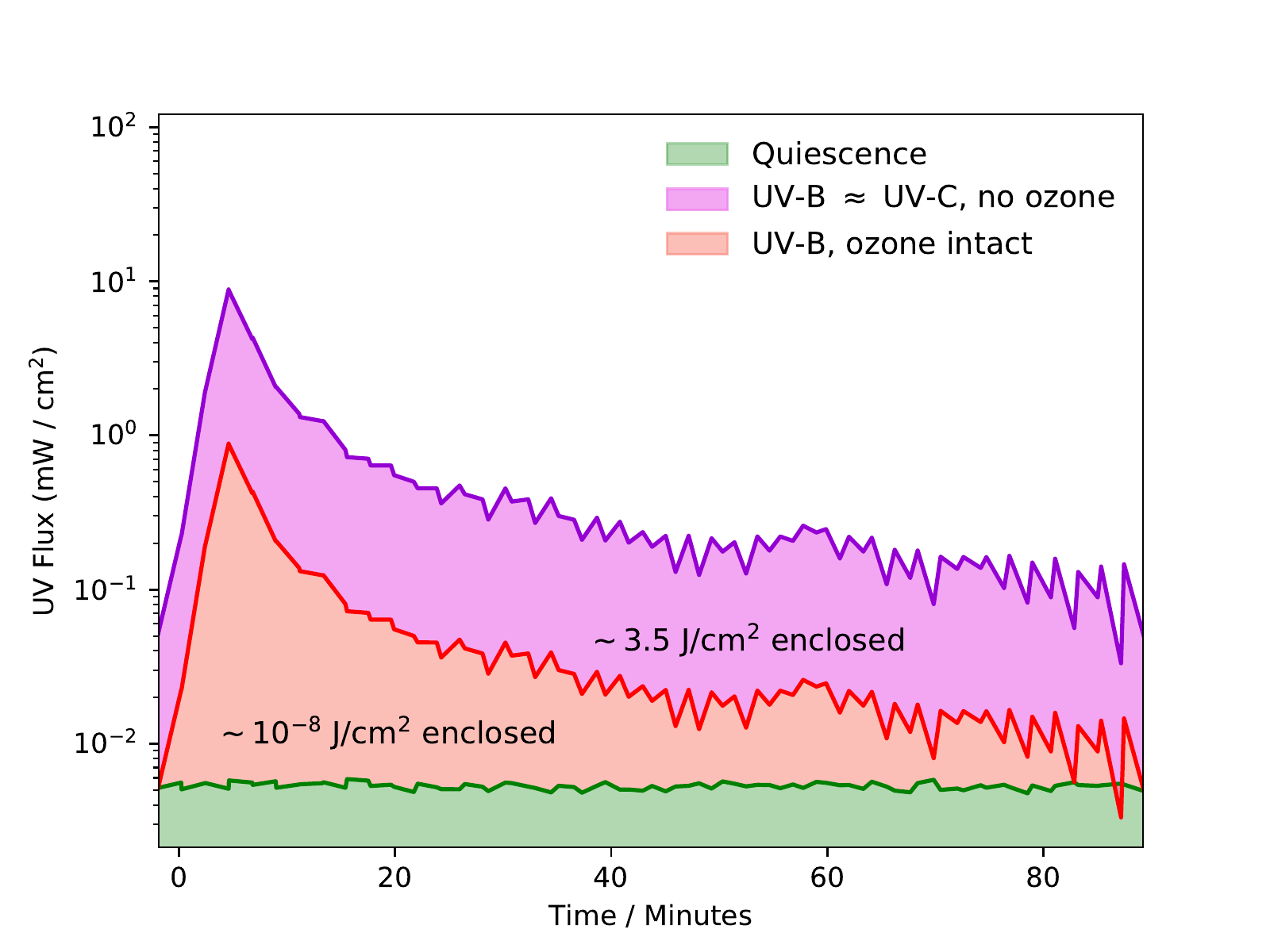}
		\label{fig:uvab_limits}
	}
	\caption{Top panel: The solid black line represents the evolution of the O$_3$ column for a planet with an Earth-like atmosphere orbiting Proxima under flares with a median proton fluence of $5.5\times10^9$ pr cm$^{-2}$. The vertical lines represent the 1- and 5- year times. The dash-dotted red line is a projection of future O$_3$ evolution. The dash-dotted projection assumes continued flare activity; the solid black line assumes no further activity after 5 years and hence returns to Earth-like conditions. The blue line is a model from the lowest end of the proton-fluence distribution (median fluence of $6.1\times10^4$ pr cm$^{-2}$) consistent with the scatter in flare and particle scaling. The brief spikes in ozone in both models and the end-of-flaring dip in the lowest-fluence model are from generation of ozone via free O atoms from photodissociation of other species. Bottom panel: Surface UV-B and UV-C flux during the superflare. Surface flux is calculated with (purple) no atmospheric attenuation from volatiles, and (red) an Earth-like, intact ozone column. UV surface levels during the superflare result in 30$\times$ lethal doses for simple microorganisms \citep{Slade2011}.}
	\label{fig:prox_}
\end{figure}

The evolution of the $\mathrm{O}_3$ column as a result of the impacting flares and proton events is shown in the top panel of Figure \ref{fig:prox_}. At the end of the simulation, 846 of 10,724 flares had generated a proton event that impacted the atmosphere of the planet, resulting in an $\mathrm{O}_3$ column loss of 90\%. The system does not appear to reach steady state with increasing time. We assess that it is likely that Proxima b has suffered extreme ozone loss. At Proxima's current activity rate, $>$99.9\% of Proxima b's $\mathrm{O}_3$ is likely to be lost within 100s of kyr, leaving the planet's surface largely unprotected from UV light. 

A complete lack of $\mathrm{O}_3$ would particularly affect the amount of germicidal UV-C reaching the surface. Although other volatiles capable of absorbing UV-C (i.e., $\mathrm{O}_2$, $\mathrm{H}_2\mathrm{O}$) are not necessarily destroyed, they do not effectively block UV-C for wavelengths longer than $\sim$2500{\AA}. No significant Earth-like atmospheric gas but $\mathrm{O}_3$ effectively absorbs in the UV-B and UV-C wavelengths $\sim$2450-2800{\AA} \citep{Tilley2017}. During the Proxima Superflare, the top-of-atmosphere receives $\sim$3.5 J cm$^{-2}$ of UV-C in the wavelength range 2400-2800 {\AA}. Absent ozone, most of this reaches the surface.

We caution that our result assumes an Earth-like atmosphere and is exploratory in nature. Ozone loss depends on indirect scaling relations between flare and particle flux, which exhibit 3-4 orders-of-magnitude of scatter (e.g. \citet{Cliver2012,Youngblood2017}). Our primary result shown in the top panel of Figure \ref{fig:prox_} assumes solar SXR-particle scaling relations \citep{Belov2005} and active M-dwarf UV-SXR scaling relations \citep{Mitra-Kraev2005}, resulting in a median proton fluence of $5.5\times10^9$ pr cm$^{-2}$. Even if we assume a $10^5$ times lower median proton fluence, a value from the lowest end of the proton-scaling distribution, we find that the ozone layer is severely depleted after 5 years. In this second run, we use the ``fiducial flare" template spectrum of Section \ref{sec:multiwavelength} and employ a planet-oriented energetic particle event probability of P = 0.25 and solar EUV-particle scaling relations and then an M dwarf synthetic spectrum \citep {Fontenla2016} to relate EUV to FUV \citep{Youngblood2017} for a median proton fluence of $6.1\times10^4$ pr cm$^{-2}$. We measure 40\% ozone loss after 5 years. While the difference between these two models is large, it is unsurprising in light of the large scatter in the correlations between solar flare intensity and proton flux. Further work is needed to more rigorously constrain the energetic particle environments of these stars.

\subsection{Effects on surface life}\label{astrobio}

Figure \ref{fig:prox_} shows the UV-B and UV-C flux at the surface of Proxima b during the superflare, given the \textit{g}\textsuperscript{$\prime$} flux measured in the Evryscope light curve, the flare spectral modeling in Section \ref{total_energy}, and assuming an orbital radius of $0.0485$ AU \citep{anglada2016}. We plot two flare regimes: (1) the expected fluxes for an intact Earth-like ozone layer (where essentially no UVC reaches the surface, and $\sim10\%$ of UVB reaches the surface; \citealt{Grant1997} and references therein); and (2) an extreme ozone-loss scenario, where UVC in the wavelength range $\sim$2450-2800{\AA} reaches the surface unimpeded by ozone or other atmospheric absorption (see Section \ref{planet_habitability}), resulting in UV-B$\simeq$UV-C surface fluxes (this equivalence is coincidental: the lack of $\sim$2500 {\AA} UV-C reaching the surface counters the proximity to the flare blackbody peak).  We assume cloudless skies.

While UV-B accounts for only 5\% of the solar UV radiation incident upon Earth, it has the largest impact upon terrestrial biology because shorter UV-C wavelengths are blocked by the Earth's atmosphere \citep{Sliney1980}. During the Proxima Superflare, 3.5 J $\mathrm{cm}^{-2}$ of UV-B reached the surface under the assumption of extreme ozone attenuation, which is below a lethal dose of 4 J $\mathrm{cm}^{-2}$ for \textit{Deinococcus radiodurans} but lethal for most UV-hardy organisms, even when protected by a shallow layer of freshwater. For example, in the top 50 cm of water, 1.5 J $\mathrm{cm}^{-2}$  of UV-B will kill 50\% of freshwater invertebrates \citep{Hurtubise1998}. Zooplankton's UV-B lethal dose is 0.5 J $\mathrm{cm}^{-2}$ \citep{Rautio2002}.

UV-C is much more efficient at damaging DNA than UV-B \citep{Rehemtulla1997,Batista2009}. Although \textit{D. radiodurans} is amongst the most UV-resilient organisms on Earth, its UV-C D90 dose (i.e., the amount of radiation required to kill 90\% of the population) of 0.0553 J $\mathrm{cm}^{-2}$ \citep{Gascon1995} is a factor of 65$\times$ smaller than the 3.6 J $\mathrm{cm}^{-2}$ reaching the surface during the Proxima Superflare, given no ozone. Recent results have suggested that more complex life such as lichens evolved for extreme environments and with adaptations such as UV-screening pigments may survive these radiation levels \citep{brandt2015}. These results suggest that surface life on areas of Proxima b exposed to these flares would have to undergo complex adaptations to survive\footnote[1]{although superflares would have a greatly reduced direct impact on organisms that may exist underground, under deep water, on the dark side of a tidally-locked world, etc.}, even if the planetary atmosphere survives the long-term impact of the stellar activity.

Earth may have undergone significantly higher UV fluxes during the early evolution of life. \citet{Rugheimer2015} give diurnally-averaged values of the surface UV-B and UV-C flux on Earth during the pre-biotic (3.9 Ga ago) and early Proterozoic (2.0 Ga ago) epochs. Assuming the full-ozone-loss scenario, the surface UV-B flux during the Proxima superflare was an average of 2$\times$ higher than that 3.9 Ga ago and a factor of 3$\times$ higher 2.0 Ga ago, although between flares the UV flux was much lower than Earth's, because late M-dwarfs are far fainter in the UV than solar-type stars. The UV-C superflare flux was a factor of 7$\times$ higher than that 3.9 Ga ago and a factor of 1750$\times$ higher than 2.0 Ga ago; again, the UV-C flux potentially reaching Proxima's surface is the critical difference compared to Earth's environment.

\section{Conclusions}\label{conclude}

Two-thirds of M-dwarfs are active \citep{west2015}, and superflares will significantly impact the habitability of the planets orbiting many of these stars, which make up the majority of the Galaxy's stellar population.  Measuring the impact of superflares on these worlds will thus be a necessary component in the search for extraterrestrial life on planets discovered by TESS \citep{Ricker_TESS} and other surveys. Beyond Proxima, Evryscope has already performed similar long-term high-cadence monitoring of every other bright Southern TESS planet-search target, and will therefore be able to measure the habitability impact of stellar activity for these stars (Howard et al., in prep.). In conjunction with coronal-mass-ejection searches from radio arrays like the LWA \citep{Hallinan2017} and Jansky Very Large Array \citep{Crosley2018}, the Evryscope will aid in constraining long-term atmospheric effects of extreme stellar activity.

On Proxima b, the superflare we detected, along with Proxima's regular and extreme activity, leads to our photochemical model predicting 90\% ozone destruction within 5 years. As Proxima's ozone column fraction does not reach a steady state at the end of that period but instead continues a clear downward trend, we conclude that Proxima b has likely suffered extreme ozone loss over long timescales. If the current activity rate of Proxima holds for longer timescales, $>$99.9\% of the planetary $\mathrm{O}_3$ is likely to be lost within 100s of kyr, leaving the planet's surface largely unprotected from UV light, and forcing extreme adaptation by any organisms on the Proxima-facing surface of Proxima b.

\section*{Acknowledgements}\label{acknowledge}
The authors wish to thank James R. Davenport, J.J. Hermes, Bart H. Dunlap, Sarah Rugheimer, and Joshua Reding for insightful comments.  WH, HC, NL, JR, CZ, and EG acknowledge funding support by the National Science Foundation CAREER grant 1555175, and the Research Corporation Scialog grants 23782 and 23822. HC is supported by the National Science Foundation Graduate Research Fellowship under Grant No. DGE-1144081. OF acknowledges funding support by the grant MDM-2014-0369 of the ICCUB (Unidad de Excelencia `Mar\'{\i}a de Maeztu'). AY acknowledges support by an appointment to the NASA Postdoctoral Program at Goddard Space Flight Center, administered by URSA through a contract with NASA. R.O.P.L. appreciates support from a STScI grant, HST-GO-14784.001-A (PI Shkolnik). E.L.S. appreciates support from NASA/Habitable Worlds grant NNX16AB62G. The Evryscope was constructed under National Science Foundation/ATI grant AST-1407589. 

{\it Facilities:} \facility{CTIO:Evryscope}, \facility{HST (STIS)}, \facility{ESO:3.6m (HARPS)}

\bibliographystyle{apj}
\bibliography{paper_references}

\begin{thebibliography}{}
\expandafter\ifx\csname natexlab\endcsname\relax\def\natexlab#1{#1}\fi

\bibitem[{{Anglada-Escud{\'e}} {et~al.}(2016){Anglada-Escud{\'e}}, {Amado},
  {Barnes}, {Berdi{\~n}as}, {Butler}, {Coleman}, {de La Cueva}, {Dreizler},
  {Endl}, {Giesers}, {Jeffers}, {Jenkins}, {Jones}, {Kiraga}, {K{\"u}rster},
  {L{\'o}pez-Gonz{\'a}lez}, {Marvin}, {Morales}, {Morin}, {Nelson}, {Ortiz},
  {Ofir}, {Paardekooper}, {Reiners}, {Rodr{\'{\i}}guez},
  {Rodr{\'{\i}}guez-L{\'o}pez}, {Sarmiento}, {Strachan}, {Tsapras}, {Tuomi}, \&
  {Zechmeister}}]{anglada2016}
{Anglada-Escud{\'e}}, G., {Amado}, P.~J., {Barnes}, J., {et~al.} 2016, \nat,
  536, 437

\bibitem[{Batista {et~al.}(2009)Batista, Kaina, Meneghini, \&
  Menck}]{Batista2009}
Batista, L.~F., Kaina, B., Meneghini, R., \& Menck, C.~F. 2009, Mutation
  Research/Reviews in Mutation Research, 681, 197

\bibitem[{{Belov} {et~al.}(2005){Belov}, {Garcia}, {Kurt}, {Mavromichalaki}, \&
  {Gerontidou}}]{Belov2005}
{Belov}, A., {Garcia}, H., {Kurt}, V., {Mavromichalaki}, H., \& {Gerontidou},
  M. 2005, \solphys, 229, 135

\bibitem[{{Bixel} \& {Apai}(2017)}]{bixel2017}
{Bixel}, A., \& {Apai}, D. 2017, \apjl, 836, L31

\bibitem[{Brandt {et~al.}(2015)Brandt, de~Vera, Onofri, \& Ott}]{brandt2015}
Brandt, A., de~Vera, J.-P., Onofri, S., \& Ott, S. 2015, International Journal
  of Astrobiology, 14, 411–425

\bibitem[{{Cinzano} {et~al.}(2001){Cinzano}, {Falchi}, \&
  {Elvidge}}]{Cinzano2001}
{Cinzano}, P., {Falchi}, F., \& {Elvidge}, C.~D. 2001, \mnras, 323, 34

\bibitem[{{Cliver} {et~al.}(2012){Cliver}, {Ling}, {Belov}, \&
  {Yashiro}}]{Cliver2012}
{Cliver}, E.~W., {Ling}, A.~G., {Belov}, A., \& {Yashiro}, S. 2012, \apjl, 756,
  L29

\bibitem[{{Crosley} \& {Osten}(2018)}]{Crosley2018}
{Crosley}, M.~K., \& {Osten}, R.~A. 2018, \apj, 856, 39

\bibitem[{{Cuntz} \& {Guinan}(2016)}]{cuntz2016}
{Cuntz}, M., \& {Guinan}, E.~F. 2016, \apj, 827, 79

\bibitem[{{Davenport} {et~al.}(2016){Davenport}, {Kipping}, {Sasselov},
  {Matthews}, \& {Cameron}}]{davenport2016}
{Davenport}, J.~R.~A., {Kipping}, D.~M., {Sasselov}, D., {Matthews}, J.~M., \&
  {Cameron}, C. 2016, \apjl, 829, L31

\bibitem[{{Davenport} {et~al.}(2014){Davenport}, {Hawley}, {Hebb},
  {Wisniewski}, {Kowalski}, {Johnson}, {Malatesta}, {Peraza}, {Keil},
  {Silverberg}, {Jansen}, {Scheffler}, {Berdis}, {Larsen}, \&
  {Hilton}}]{davenport2014}
{Davenport}, J.~R.~A., {Hawley}, S.~L., {Hebb}, L., {et~al.} 2014, \apj, 797,
  122

\bibitem[{{Fontenla} {et~al.}(2016){Fontenla}, {Linsky}, {Witbrod}, {France},
  {Buccino}, {Mauas}, {Vieytes}, \& {Walkowicz}}]{Fontenla2016}
{Fontenla}, J.~M., {Linsky}, J.~L., {Witbrod}, J., {et~al.} 2016, \apj, 830,
  154

\bibitem[{Gasc{\'o}n {et~al.}(1995)Gasc{\'o}n, Oubi{\~{n}}a, P{\'e}rez-Lezaun,
  \& Urmeneta}]{Gascon1995}
Gasc{\'o}n, J., Oubi{\~{n}}a, A., P{\'e}rez-Lezaun, A., \& Urmeneta, J. 1995,
  Current Microbiology, 30, 177

\bibitem[{{Grant} \& {Heisler}(1997)}]{Grant1997}
{Grant}, R.~H., \& {Heisler}, M. 1997, Journal of Applied Meteorology, 36, 1336

\bibitem[{{Grie{\ss}meier} {et~al.}(2009){Grie{\ss}meier}, {Stadelmann},
  {Grenfell}, {Lammer}, \& {Motschmann}}]{griessmeier2009}
{Grie{\ss}meier}, J.-M., {Stadelmann}, A., {Grenfell}, J.~L., {Lammer}, H., \&
  {Motschmann}, U. 2009, \icarus, 199, 526

\bibitem[{{Grie{\ss}meier} {et~al.}(2016){Grie{\ss}meier}, {Tabataba-Vakili},
  {Stadelmann}, {Grenfell}, \& {Atri}}]{Greissmeier2016}
{Grie{\ss}meier}, J.-M., {Tabataba-Vakili}, F., {Stadelmann}, A., {Grenfell},
  J.~L., \& {Atri}, D. 2016, \aap, 587, A159

\bibitem[{{G{\"u}del} {et~al.}(2004){G{\"u}del}, {Audard}, {Reale}, {Skinner},
  \& {Linsky}}]{gudel2004}
{G{\"u}del}, M., {Audard}, M., {Reale}, F., {Skinner}, S.~L., \& {Linsky},
  J.~L. 2004, \aap, 416, 713

\bibitem[{{Hallinan} \& {Anderson}(2017)}]{Hallinan2017}
{Hallinan}, G., \& {Anderson}, M.~M. 2017, in Radio Exploration of Planetary
  Habitability (AASTCS5), Vol.~49, 401.01

\bibitem[{{Hawley} {et~al.}(2014){Hawley}, {Davenport}, {Kowalski},
  {Wisniewski}, {Hebb}, {Deitrick}, \& {Hilton}}]{hawley2014}
{Hawley}, S.~L., {Davenport}, J.~R.~A., {Kowalski}, A.~F., {et~al.} 2014, \apj,
  797, 121

\bibitem[{{Hurtubise} {et~al.}(1998){Hurtubise}, {Havel}, \&
  {Little}}]{Hurtubise1998}
{Hurtubise}, R.~D., {Havel}, J.~E., \& {Little}, E.~E. 1998, Limnology and
  Oceanography, 43, 1082

\bibitem[{{Kowalski} {et~al.}(2010){Kowalski}, {Hawley}, {Holtzman},
  {Wisniewski}, \& {Hilton}}]{kowalski2010}
{Kowalski}, A.~F., {Hawley}, S.~L., {Holtzman}, J.~A., {Wisniewski}, J.~P., \&
  {Hilton}, E.~J. 2010, \apjl, 714, L98

\bibitem[{{Law} {et~al.}(2016){Law}, {Fors}, {Ratzloff}, {Corbett}, {del Ser},
  \& {Wulfken}}]{Law2016}
{Law}, N.~M., {Fors}, O., {Ratzloff}, J., {et~al.} 2016, in \procspie, Vol.
  9906, Ground-based and Airborne Telescopes VI, 99061M

\bibitem[{{Law} {et~al.}(2015){Law}, {Fors}, {Ratzloff}, {Wulfken},
  {Kavanaugh}, {Sitar}, {Pruett}, {Birchard}, {Barlow}, {Cannon}, {Cenko},
  {Dunlap}, {Kraus}, \& {Maccarone}}]{Evryscope2015}
{Law}, N.~M., {Fors}, O., {Ratzloff}, J., {et~al.} 2015, \pasp, 127, 234

\bibitem[{{Liang} {et~al.}(2016){Liang}, {Wang}, {Zhou}, {Zhou}, {Zhang},
  {Xie}, {Liu}, {Wang}, \& {Ashley}}]{Liang2016}
{Liang}, E.-S., {Wang}, S., {Zhou}, J.-L., {et~al.} 2016, \aj, 152, 168

\bibitem[{{Loyd} {et~al.}(in prep.)}]{loyd2018}
{Loyd}, R. O.~P., {et~al.} in prep., \apj

\bibitem[{{Lurie} {et~al.}(2014){Lurie}, {Henry}, {Jao}, {Quinn}, {Winters},
  {Ianna}, {Koerner}, {Riedel}, \& {Subasavage}}]{lurie2014}
{Lurie}, J.~C., {Henry}, T.~J., {Jao}, W.-C., {et~al.} 2014, \aj, 148, 91

\bibitem[{{MacGregor} {et~al.}(2018){MacGregor}, {Weinberger}, {Wilner},
  {Kowalski}, \& {Cranmer}}]{macgregor2018}
{MacGregor}, M.~A., {Weinberger}, A.~J., {Wilner}, D.~J., {Kowalski}, A.~F., \&
  {Cranmer}, S.~R. 2018, \apjl, 855, L2

\bibitem[{{Mayor} {et~al.}(2003){Mayor}, {Pepe}, {Queloz}, {Bouchy},
  {Rupprecht}, {Lo Curto}, {Avila}, {Benz}, {Bertaux}, {Bonfils}, {Dall},
  {Dekker}, {Delabre}, {Eckert}, {Fleury}, {Gilliotte}, {Gojak}, {Guzman},
  {Kohler}, {Lizon}, {Longinotti}, {Lovis}, {Megevand}, {Pasquini}, {Reyes},
  {Sivan}, {Sosnowska}, {Soto}, {Udry}, {van Kesteren}, {Weber}, \&
  {Weilenmann}}]{Mayor2003}
{Mayor}, M., {Pepe}, F., {Queloz}, D., {et~al.} 2003, The Messenger, 114, 20

\bibitem[{{Meadows} {et~al.}(2018){Meadows}, N., W., Jacob, P., Tyler, D.,
  Russell, K., P., Rodrigo, E., R., \& David}]{Meadows2018}
{Meadows}, V.~S., N., A.~G., W., S.~E., {et~al.} 2018, Astrobiology, 18, 133,
  pMID: 29431479

\bibitem[{{Mitra-Kraev} {et~al.}(2005){Mitra-Kraev}, {Harra}, {G{\"u}del},
  {Audard}, {Branduardi-Raymont}, {Kay}, {Mewe}, {Raassen}, \& {van
  Driel-Gesztelyi}}]{Mitra-Kraev2005}
{Mitra-Kraev}, U., {Harra}, L.~K., {G{\"u}del}, M., {et~al.} 2005, \aap, 431,
  679

\bibitem[{{Osten} \& {Wolk}(2015)}]{Osten2015}
{Osten}, R.~A., \& {Wolk}, S.~J. 2015, \apj, 809, 79

\bibitem[{{Owen} \& {Mohanty}(2016)}]{owen2016}
{Owen}, J.~E., \& {Mohanty}, S. 2016, \mnras, 459, 4088

\bibitem[{{Pavlenko} {et~al.}(2017){Pavlenko}, {Su{\'a}rez Mascare{\~n}o},
  {Rebolo}, {Lodieu}, {B{\'e}jar}, \& {Gonz{\'a}lez
  Hern{\'a}ndez}}]{Pavlenko2017}
{Pavlenko}, Y., {Su{\'a}rez Mascare{\~n}o}, A., {Rebolo}, R., {et~al.} 2017,
  \aap, 606, A49

\bibitem[{Rautio \& Korhola(2002)}]{Rautio2002}
Rautio, M., \& Korhola, A. 2002, Polar Biology, 25, 460

\bibitem[{Rehemtulla {et~al.}(1997)Rehemtulla, Hamilton, Chinnaiyan, \&
  Dixit}]{Rehemtulla1997}
Rehemtulla, A., Hamilton, C.~A., Chinnaiyan, A.~M., \& Dixit, V.~M. 1997, The
  Journal of biological chemistry, 272, 25783

\bibitem[{{Ribas} {et~al.}(2016){Ribas}, {Bolmont}, {Selsis}, {Reiners},
  {Leconte}, {Raymond}, {Engle}, {Guinan}, {Morin}, {Turbet}, {Forget}, \&
  {Anglada-Escud{\'e}}}]{Ribas2016}
{Ribas}, I., {Bolmont}, E., {Selsis}, F., {et~al.} 2016, \aap, 596, A111

\bibitem[{{Ricker} {et~al.}(2014){Ricker}, {Winn}, {Vanderspek}, {Latham},
  {Bakos}, {Bean}, {Berta-Thompson}, {Brown}, {Buchhave}, {Butler}, {Butler},
  {Chaplin}, {Charbonneau}, {Christensen-Dalsgaard}, {Clampin}, {Deming},
  {Doty}, {De Lee}, {Dressing}, {Dunham}, {Endl}, {Fressin}, {Ge}, {Henning},
  {Holman}, {Howard}, {Ida}, {Jenkins}, {Jernigan}, {Johnson}, {Kaltenegger},
  {Kawai}, {Kjeldsen}, {Laughlin}, {Levine}, {Lin}, {Lissauer}, {MacQueen},
  {Marcy}, {McCullough}, {Morton}, {Narita}, {Paegert}, {Palle}, {Pepe},
  {Pepper}, {Quirrenbach}, {Rinehart}, {Sasselov}, {Sato}, {Seager},
  {Sozzetti}, {Stassun}, {Sullivan}, {Szentgyorgyi}, {Torres}, {Udry}, \&
  {Villasenor}}]{Ricker_TESS}
{Ricker}, G.~R., {Winn}, J.~N., {Vanderspek}, R., {et~al.} 2014, in \procspie,
  Vol. 9143, Space Telescopes and Instrumentation 2014: Optical, Infrared, and
  Millimeter Wave, 914320

\bibitem[{{Rugheimer} {et~al.}(2015){Rugheimer}, {Kaltenegger}, {Segura},
  {Linsky}, \& {Mohanty}}]{Rugheimer2015}
{Rugheimer}, S., {Kaltenegger}, L., {Segura}, A., {Linsky}, J., \& {Mohanty},
  S. 2015, \apj, 809, 57

\bibitem[{{Scalo} {et~al.}(2007){Scalo}, {Kaltenegger}, {Segura}, {Fridlund},
  {Ribas}, {Kulikov}, {Grenfell}, {Rauer}, {Odert}, {Leitzinger}, {Selsis},
  {Khodachenko}, {Eiroa}, {Kasting}, \& {Lammer}}]{scalo2007}
{Scalo}, J., {Kaltenegger}, L., {Segura}, A.~G., {et~al.} 2007, Astrobiology,
  7, 85

\bibitem[{{Schaefer}(1990)}]{Schaefer1990}
{Schaefer}, B.~E. 1990, \pasp, 102, 212

\bibitem[{{Seager} \& {Deming}(2010)}]{seager2010}
{Seager}, S., \& {Deming}, D. 2010, \araa, 48, 631

\bibitem[{{Segura} {et~al.}(2010){Segura}, {Walkowicz}, {Meadows}, {Kasting},
  \& {Hawley}}]{Segura2010}
{Segura}, A., {Walkowicz}, L.~M., {Meadows}, V., {Kasting}, J., \& {Hawley}, S.
  2010, Astrobiology, 10, 751

\bibitem[{{Shields} {et~al.}(2016){Shields}, {Ballard}, \&
  {Johnson}}]{shields2016}
{Shields}, A.~L., {Ballard}, S., \& {Johnson}, J.~A. 2016, ArXiv e-prints,
  arXiv:1610.05765

\bibitem[{Slade \& Radman(2011)}]{Slade2011}
Slade, D., \& Radman, M. 2011, Microbiology and Molecular Biology Reviews, 75,
  133

\bibitem[{Sliney \& Wolbarsht(1980)}]{Sliney1980}
Sliney, D., \& Wolbarsht, M. 1980, in Safety with lasers and other optical
  sources, Vol.~. (Plenum Press), 194

\bibitem[{{Tabataba-Vakili} {et~al.}(2016){Tabataba-Vakili}, {Grenfell},
  {Grie{\ss}meier}, \& {Rauer}}]{Tabataba2016}
{Tabataba-Vakili}, F., {Grenfell}, J.~L., {Grie{\ss}meier}, J.-M., \& {Rauer},
  H. 2016, \aap, 585, A96

\bibitem[{{Tamuz} {et~al.}(2005){Tamuz}, {Mazeh}, \& {Zucker}}]{tamuz2005}
{Tamuz}, O., {Mazeh}, T., \& {Zucker}, S. 2005, \mnras, 356, 1466

\bibitem[{{Tarter} {et~al.}(2007){Tarter}, {Backus}, {Mancinelli}, {Aurnou},
  {Backman}, {Basri}, {Boss}, {Clarke}, {Deming}, {Doyle}, {Feigelson},
  {Freund}, {Grinspoon}, {Haberle}, {Hauck}, {Heath}, {Henry}, {Hollingsworth},
  {Joshi}, {Kilston}, {Liu}, {Meikle}, {Reid}, {Rothschild}, {Scalo}, {Segura},
  {Tang}, {Tiedje}, {Turnbull}, {Walkowicz}, {Weber}, \& {Young}}]{tarter2007}
{Tarter}, J.~C., {Backus}, P.~R., {Mancinelli}, R.~L., {et~al.} 2007,
  Astrobiology, 7, 30

\bibitem[{{Tilley} {et~al.}(2017){Tilley}, {Segura}, {Meadows}, {Hawley}, \&
  {Davenport}}]{Tilley2017}
{Tilley}, M.~A., {Segura}, A., {Meadows}, V.~S., {Hawley}, S., \& {Davenport},
  J. 2017, ArXiv e-prints, arXiv:1711.08484

\bibitem[{{Walker}(1981)}]{Walker1981}
{Walker}, A.~R. 1981, \mnras, 195, 1029

\bibitem[{{Walker} {et~al.}(2003){Walker}, {Matthews}, {Kuschnig}, {Johnson},
  {Rucinski}, {Pazder}, {Burley}, {Walker}, {Skaret}, {Zee}, {Grocott},
  {Carroll}, {Sinclair}, {Sturgeon}, \& {Harron}}]{Walker2003}
{Walker}, G., {Matthews}, J., {Kuschnig}, R., {et~al.} 2003, \pasp, 115, 1023

\bibitem[{{Weaver}(1947)}]{Weaver1947}
{Weaver}, H.~F. 1947, \pasp, 59, 232

\bibitem[{{West} {et~al.}(2015){West}, {Weisenburger}, {Irwin},
  {Berta-Thompson}, {Charbonneau}, {Dittmann}, \& {Pineda}}]{west2015}
{West}, A.~A., {Weisenburger}, K.~L., {Irwin}, J., {et~al.} 2015, \apj, 812, 3

\bibitem[{{Yashiro} {et~al.}(2006){Yashiro}, {Akiyama}, {Gopalswamy}, \&
  {Howard}}]{Yashiro2006}
{Yashiro}, S., {Akiyama}, S., {Gopalswamy}, N., \& {Howard}, R.~A. 2006, \apj,
  650, L143

\bibitem[{{Youngblood} {et~al.}(2017){Youngblood}, {France}, {Loyd}, {Brown},
  {Mason}, {Schneider}, {Tilley}, {Berta-Thompson}, {Buccino}, {Froning},
  {Hawley}, {Linsky}, {Mauas}, {Redfield}, {Kowalski}, {Miguel}, {Newton},
  {Rugheimer}, {Segura}, {Roberge}, \& {Vieytes}}]{Youngblood2017}
{Youngblood}, A., {France}, K., {Loyd}, R.~O.~P., {et~al.} 2017, \apj, 843, 31

\end{thebibliography}

\end{document}